\documentclass[10pt,letterpaper,onecolumn,oneside]{article}
\usepackage[top=1in,left=1.25in,footskip=0.75in,marginparwidth=2in]{geometry}

\usepackage{amsmath}
\usepackage{booktabs} 
\usepackage{nopageno}
\usepackage{rotating}

\usepackage[utf8]{inputenc}

\usepackage[right]{lineno}

\usepackage{microtype}
\DisableLigatures[f]{encoding = *, family = * }

\usepackage{changepage}

\usepackage[aboveskip=1pt,labelfont=bf,labelsep=period,singlelinecheck=off]{caption}

\makeatletter
\renewcommand{\@biblabel}[1]{\quad#1.}
\makeatother

\usepackage{graphicx}
\usepackage{amssymb}
\usepackage{epstopdf}
\usepackage{comment}
\usepackage{enumerate}
\usepackage{color}
\usepackage{subeqnarray}
\usepackage{ulem}

\usepackage{lastpage,fancyhdr,graphicx}
\usepackage{epstopdf}
\usepackage{color}

\definecolor{Gray}{gray}{.25}

\usepackage{graphicx}

\usepackage{sidecap}

\usepackage{wrapfig}
\usepackage[pscoord]{eso-pic}
\usepackage[fulladjust]{marginnote}
\reversemarginpar

\usepackage{comment}
\usepackage{siunitx}
\usepackage{textcomp}
\usepackage{units}
\usepackage{ulem}

\usepackage{epstopdf, epsfig}
\usepackage[english]{babel}
\usepackage{siunitx}
\usepackage{comment}
\usepackage{amsmath} 
\usepackage{breqn}

\usepackage{graphics} 
\usepackage{epsfig} 

\begin{document}

\thispagestyle{plain}

\vspace*{0.35in}

\begin{flushleft}
{\Large
\textbf\newline{Simultaneous PIV and LIF measurements in stratified flows using pulsed lasers}
}
\newline
\\
{Paolo Luzzatto-Fegiz}\footnote{ pfegiz@ucsb.edu}

\bigskip
Department of Mechanical Engineering, University of California, Santa Barbara, 93106, USA

\end{flushleft}

\begin{abstract}
We examine the problem of performing simultaneous and coplanar Particle Image Velocimetry (PIV) and Laser-Induced Fluorescence (LIF) measurements in a stratified fluid initially at rest. Our focus is on enabling detailed velocity and density measurements in long internal waves and gravity currents, through relatively small modifications of typical existing PIV systems comprising pulsed lasers, using dye concentration as a proxy for fluid density.
Several issues have limited such measurements. These include: (1) variations in the laser intensity and beam structure between laser pulses; (2) PIV particles concentrating preferentially at their neutral buoyancy depth, thereby yielding nonuniform dye illumination; and (3) the need to maintain a constant index of refraction across large stratified fluid volumes. These issues can cause large errors in the measured dye concentration, and therefore in the inferred density.
Here we focus on an experimental setup comprising a stratified layer overlaying a deep homogeneous region. We produce long internal waves using a lock-release setup, in order to investigate the structure of waves comprising recirculating fluid regions (known as ``trapped cores''), which are of current interest in oceanographic applications. 
We maintain a short optical path and use velocity information from PIV data to minimize index-of-refraction issues. We exploit the fact that the system is initially at rest to devise a mapping that links apparent and actual dye concentration, thus sidestepping nonuniform illumination issues due to particle clustering. Finally, we devise a procedure to correct for laser power variations along each ray in the sheet.
We test our technique for a small-amplitude wave, where the density can also be separately calculated  without resorting to LIF, finding good agreement. Furthermore, we show results for large-amplitude waves, where our approach yields the first measurements of the density field in long internal waves with trapped cores, thereby providing detailed data for the development of theoretical models.
\end{abstract}

Particle Image Velocimetry (PIV) and Laser Induced Fluorescence (LIF) can provide complementary information about a flow, as they may enable tracking simultaneously the evolution of the velocity field, as well as the behaviour of a scalar. Early experiments combing PIV and LIF appear to have focused primarily on uniform-density flows (e.g.~\cite{Simoens_Ayrault_EF_1994}), with the scalar behaving as a passive tracer. When considering density-stratified flows, the scalar can in principle be used as a proxy for the fluid density, by defining a suitable calibration. However, the introduction of density gradients brings several new issues, which must be resolved in order to perform accurate measurements of the scalar field.

\begin{figure}[t] 
	\centerline{
		\includegraphics{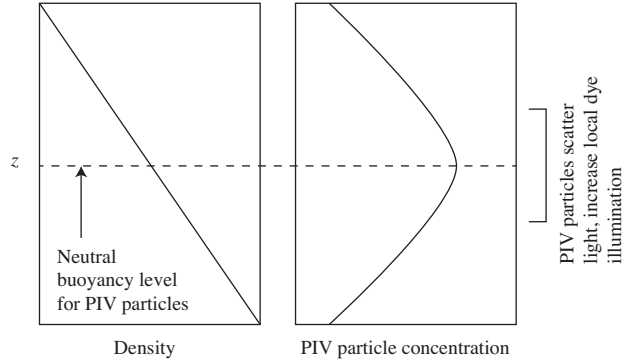}}
	\caption{Sketch of the PIV particle concentration in a stratified fluid, leading to enhanced illumination near the depth at which the particles are neutrally buoyant.}
	\label{fig:particleSketch}
\end{figure}

A first issue involves optical distortions associated with density gradients (for example, if a straightforward salt stratification is used), since the resulting scalar measurements are affected in two ways. Firstly, errors are introduced throughout the imaged field, as the optical path through which each fluid particle is imaged depends on time; this error can be estimated through a ray-tracing exercise, and can often be limited by maintaining a short optical path. A second problem can arise when sharp spatial gradients in refractive index can reshape the light pattern used for illumination. When vortex shedding or other relatively small-scale features are involved, this results in characteristic `streaks' appearing in the scalar field \cite{Petracci_etal_2006}.
Several experiments have made use of refractive-index-matching techniques to minimize these optical distortions (for example, by using glycerol and potassium phosphate \cite{ALAHYARI:1994vz} or salt and isopropyl alcohol \cite{Troy_Koseff_EF_2005}); however, index matching becomes more expensive and technically challenging as the scale of the facility increases.
\cite{Petracci_etal_2006} sidestepped the need for index matching by maintaining relatively short optical paths, and applying a Fourier filter to their scalar fields, thereby effectively removing the streaks. However, this came at the cost of also removing some of the smaller features in the scalar field.

A second issue, when performing simultaneous PIV and LIF in stratified fluids, is associated with the tendency of PIV particles to concentrate preferentially near the depth where they would be neutrally buoyant, as sketched in figure~\ref{fig:particleSketch}. Since the particles scatter light, this leads to stronger effective illumination in regions of higher particle density, resulting in a locally higher dye fluorescence, and therefore in greater apparent dye concentration, thereby introducing an error into the measurement. Note that even a camera with a filter that would pass only the fluorescence wavelength would be affected by this error. This is a consequence of the fact that the additional scattered light is absorbed by the dye, resulting in greater fluorescence. Therefore the camera receives an increased amount of light that follows the expected fluorescence spectrum of the dye. By contrast, any issues associated with light that reaches the camera {\it directly} after being scattered by the particles could be removed by a suitable filter.

\cite{Law_Wang_ETFS_2000} addressed this issue by introducing an additional calibration, which essentially links particle concentration to enhancement in illumination. Implementing this technique requires measuring the particle concentration throughout the experiment and establishing this additional calibration. Furthermore, stratified flow experiments often use particles that are not monodisperse, to help ensure that the full range of depths are seeded (since different-sized particles can retain different amounts of gas on their surface, leading to a broader effective range of specific gravities; see e.g.~\cite{Carr_etal_PF_2015}). This would introduce, as an additional complication, the need to account also for particle size in Law~\&~Wang's calibration.

\begin{figure}[t] 
	\centerline{
		\includegraphics{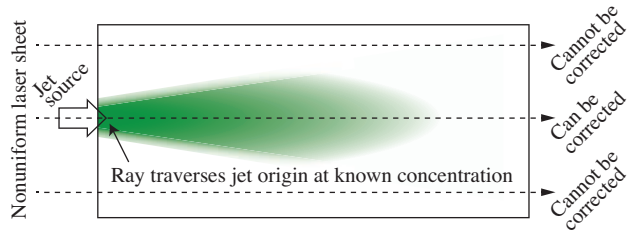}}
	\caption{Sketch of the correction, introduced in \cite{Guillard_etal_EF_1998}, to account for time-dependent nonuniformities in the laser sheet. The approach was developed for a dyed jet, flowing into a clear ambient. Only rays traversing regions of known concentration (here, near the jet source) can be corrected.}
	\label{fig:GuillardSketch}
\end{figure}

Choosing how to illuminate the particles and dye has also been a source of complications when performing combined PIV and LIF. Pulsed lasers are often used to perform PIV measurements; however, they can give significant variations in power output between shots, resulting in inaccurate dye concentration measurements \cite{Martin_Garcia_EF_2008}. On the other hand, continuous lasers offer more stable output, but their lower intensity might limit the resolution in PIV measurements. A comprehensive and effective solution involves combining two lasers into a single optical path, and switching between the two (e.g.~\cite{Fukushima_etal_Springer_2002}). While very effective, this solution brings substantial additional complexity; furthermore, it does not address the two issues described earlier, involving index of refraction and increased dye illumination. 
Another (possibly practically simpler, as well as cheaper) choice of illumination involves using only one laser. A continuous-beam laser offers best performance for LIF measurements (while placing some limits on the resolution of the velocity measurements). On the other hand, since pulsed lasers are often used in PIV setups, it is interesting to look for a relatively simple extension to also enable accurate LIF with these systems. One can meter the variation in output between pulses, and correct the LIF data accordingly \cite{Martin_Garcia_EF_2008}. However, shot-to-shot variations are not necessarily limited to intensity, but can also affect beam structure, resulting in time-dependent nonuniformities in the resulting light sheet. In principle, one could account for these spatial variations by separately applying a correction to each ray composing the sheet. However, one must be able to calculate the correction for each ray. This can be achieved for rays that are traversing regions where the actual dye concentration is known at all times. \cite{Guillard_etal_EF_1998} developed this approach for a dyed jet flowing into a clear ambient fluid, as sketched in figure~\ref{fig:GuillardSketch}. In this example, the jet source is at a known concentration. Note that a persistent limit of this approach is that rays that do not traverse any regions of known concentration cannot be corrected (as shown in figure~\ref{fig:GuillardSketch}).


In summary, the issues discussed above, which affect simultaneous PIV and LIF measurements in stratified fluids, using pulsed lasers, are:

\begin{itemize}
	\item Time- and space-dependent variations in the light sheet;
	\item Preferential clustering of PIV particles near their neutral buoyancy depth, leading to locally higher dye fluorescence; and
	\item Index-of-refraction variations, affecting both illumination and data acquisition.
\end{itemize}

\begin{figure*}[t]
	\centerline{
		\includegraphics{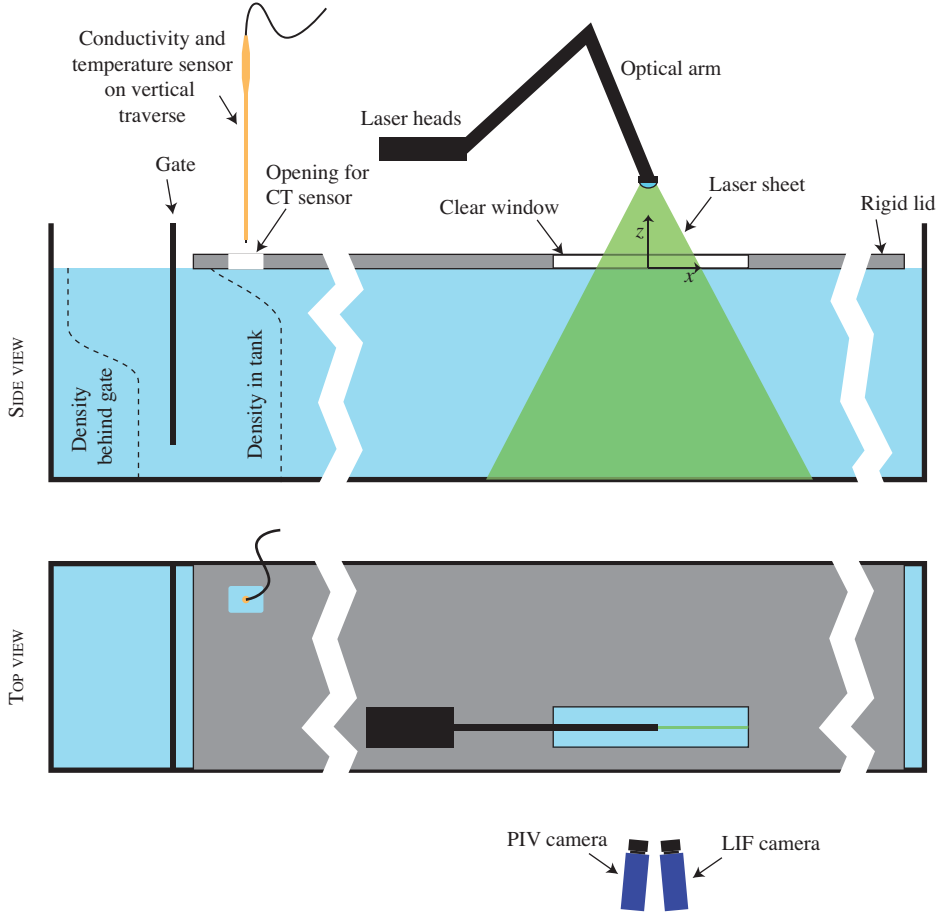}}
	\caption{Sketch of the overall experimental setup (not to scale).}
	\label{fig:tankSketch}
\end{figure*}

In this paper, we develop solutions to address each of these issues, for an experiment investigating the structure of long internal waves. In section~\ref{sec:setup} we describe our experimental setup. Section~\ref{sec:power} explains how we address power fluctuations, whereas section~\ref{sec:scatter} describes how we minimize errors due to scatter from PIV particles, as well as from index-of-refraction variations. Section~\ref{sec:example} shows examples involving internal solitary wave measurements. Finally, in section \ref{sec:conclusions} we draw conclusions from the use of the overall technique.


\section{Experimental Setup\label{sec:setup}}
The overall tank length, width and depth are 8\,m $\times$ $0.4\,$m $\times$ 0.5\,m, respectively (a sketch of the setup is shown in figure~\ref{fig:tankSketch}). In all experiments, a deep and uniform layer of saltwater was covered with a shallow stratified layer of height $h$, yielding an overall depth $H$. An approximately linear stratification was produced using the two-bucket method \cite{Economidou_Hunt_EF_2008}, using mixtures of freshwater and filtered seawater. The background profile was measured using a conductivity-temperature probe from Precision Measurement Engineering \cite{Head_PhD_1983}; density was calculated using the Seawater library, version 3.3.

Immersion pumps were used to help remove dissolved gases from the water reservoirs, to minimize opportunities for bubble formation on the walls or on the conductivity probe. Before each experiment, we waited for the water to be in thermal equilibrium with the room (the immersion pumps proved useful in initially raising the temperature of the water closer to that of the room). 

In all experiments reported here, $h/H \simeq 0.15$, whereas the total depth $H \simeq 0.35$\,m. The internal buoyancy frequency $N$, defined by
\begin{equation}
N^2 =  -\frac{g}{\rho} \frac{\partial \rho}{\partial z} 
\end{equation}
had a typical value, in the stratified layer, of $N_0\simeq 1.5$\,s$^{-1}$ (here $g$ is the acceleration due to gravity, $\rho$ is the density, and $z$ is the vertical (upward pointing) direction.
While experimental Reynolds numbers in the literature are often based on overall depth $H$, for small $h/H$ the most important length scale is arguably $h$. For a wave with phase velocity $c$, we therefore define a Reynolds number $Re_h = ch / \nu$, where $\nu$ is the kinematic viscosity. In our experiments, $Re_h \simeq 4,000$.
Waves are produced using a classic lock-release arrangement \cite{Grue_etal_JFM_2000,Sutherland_etal_JGR-O_2013} at one end of the tank. Once the background stratification has been set up, a gate is introduced (approximately 0.3\,m from one end of the tank), and additional liquid (closely matching the surface density) is added. Removing the gate generates a surface-propagating gravity current, whose front radiates long internal waves.

We image two downstream regions, located approximately 3.2\,m and 6.2\,m from the end of the tank at which the lock-release mechanism is located. We acquire velocity data in a single plane by means of a particle image velocimetry (PIV) setup, comprising an Nd:Yag pulsed laser (operating at 532\,nm) and CCD cameras with resolution $2048 \times 2048$ (LaVision). The field of view is approximately 0.35\,m wide, and is located in a plane that is about $0.07$\,m away from the side of the tank, to minimize side-wall effects, while maintaining a relatively short optical path to reduce optical distortion. The flow is seeded with Vestosint particles with a nominal size of 15\,$\mu$m; a relatively high seeding density is used, 
thus enabling calculation of independent velocity vectors over a length scale of about $3\times10^{-3}$m. The typical data acquisition rate is 7\,Hz; velocity fields are calculated using DaVis~7.2.

To also measure density fields that are co-planar with the velocity data, we performed simultaneous planar laser-induced fluorescence (LIF), using Rhodamine WT as the fluorescent dye. The dye is added to either the saltwater or freshwater reservoir before the experiment (this choice is guided by the analysis developed in section~\ref{sec:scatter}). In order to neutralize any chlorine in the freshwater (which would bleach the dye), we introduce 12\,PPM (by weight) of ascorbic acid crystals in the freshwater reservoir. 
The PIV camera is equipped with a notch filter, together with a polarized filter to reduce light from reflections. The LIF camera uses a 570\,nm high-pass filter to avoid capturing any light directly scattered by the PIV particles. Following \cite{Melton_Lipp_EF_2003}, we warm up the lasers for approximately 20 minutes before each experiment. The power setting for the laser was determined through preliminary experiments to ensure that, in the field of view, the fluorescent response was linear with respect to dye concentration (at least up to a maximum concentration of 10\,PPB).

Recent work has shown that, in the presence of a free surface, solitary internal waves can be drastically affected by Marangoni effects developing at the air-water interface \cite{Luzzatto-Fegiz_Helfrich_JFM_2014}. For this reason, we perform experiments with a no-slip boundary at the top, which is introduced shortly after the tank has been filled, using styrofoam panels whose width matches precisely that of the tank. We were careful to ensure that the panels had no seams or joins at least 1\,m upstream and downstream of each imaging window, and did not have appreciable gaps elsewhere (aside from the ends of the tank). A small opening was cut for lowering the conductivity-temperature probe. The panels were lowered carefully until fully wet, and were held in place rigidly. In order to enhance wetting and minimize bubbles, we treated the panels with a weak solution of dechlorinated water and wetting agent (0.2\% Photo-Flo by volume), which was allowed to dry before the panels were lowered.

Once a set of images was acquired, and the laboratory-frame velocities $(u,w)$ had been calculated, the phase velocity $c$ was estimated by tracking the $x$-location of the wave trough. To achieve this, we followed the horizontal motion, at fixed depth, of the point at which $w(x, z = z_0) = 0$, where $z_0$ is the depth corresponding to the lower edge of the stratified layer. The uncertainty in $c$ is estimated (by bootstrapping) to be less than 1\%. The amplitude $\eta_{\mathrm{pycno}}$ is defined as the displacement of the isopycnal at the lower end of the stratification (relative to initial condition), and is measured either by first computing streamlines in the moving frame (for small-amplitude waves, without instabilities or trapped cores) or by tracing the appropriate isopycnal in the density field. 
To obtain an overall view of the waves, we construct composite views, by shifting successive images horizontally by a distance $c\Delta t$ (where $\Delta t$ is the time between images), interpolating to a common grid, and averaging data from overlapping image regions.

\section{Performing simultaneous PIV and LIF measurements in stratified flows using pulsed lasers}
\subsection{Accounting for nonuniform power fluctuations\label{sec:power}}
\begin{figure}
\centering
\includegraphics[width=0.5\linewidth]{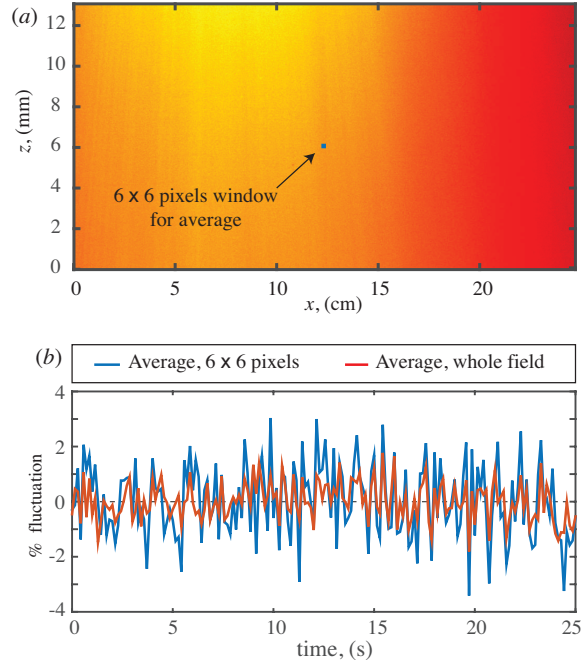}
\caption{Fluctuation in laser power over time. ($a$) Measured fluorescence for a sample of uniform 10\,PPB dye concentration. ($b$) time traces for average measurements over a $6\times6$ pixels window (blue) and over the whole field (red).}
\label{fig:powerFluctuation_sideBySide}
\end{figure}

In this section, we consider the effect of power fluctuations, in the absence of PIV particles. Figure~\ref{fig:powerFluctuation_sideBySide}($a$) shows an image taken with a sample of uniform density and dye concentration. In figure~\ref{fig:powerFluctuation_sideBySide}($b$), we plot the fluctuation in the measured fluorescence, for a small region of $6\times6$ pixels (in blue), as well as for the whole field of view (in red). While local and global fluctuations exhibit some correlation, they can also be markedly different. This shows the need for calculating and applying a {\it local} correction (that is, one that can be applied along each ray). 

In order to avoid having to correct for dye absorption, we employ relatively low maximum dye concentrations (approximately 10 PPB). If the laser power is steady, the measured intensity $I$ at each CCD pixel (with horizontal and vertical indices $i,j$)  is related to the concentration field $C$ for each image $n$ by \cite{Crimaldi_EF_2008}
\begin{equation}
	I_n(i,j) = \alpha(i,j) \, C_n(i,j) + I_0(i,j),
\end{equation}
where $\alpha(i,j)$ is an empirically constructed calibration matrix, and $I_0(i,j)$ is the background level for each pixel (which can be acquired by taking a series of images with the lens cap on). If the illumination is time-dependent, such that the laser intensity changes with each image $n$, we write
\begin{equation}\label{eq:IwithAlpha}
	I_n(i,j) = \alpha(i,j) \, a_n(\theta)  \, C_n(i,j) + I_0(i,j),
\end{equation}
where $a_n(\theta)$ is a (time-dependent) factor capturing the power variation with azimuthal angle $\theta$ across the sheet. Without loss of generality, we define $a_n(\theta)$ such that its long-term average is~1, that is
\begin{equation}
\bar{a} (\theta) = \lim_{N\rightarrow\infty} \frac{1}{N}\sum_{n=1}^N a_n(\theta) \equiv 1,
\end{equation}
 Therefore we evaluate $\alpha(i,j)$ by taking $N$ images at a known reference concentration $C(i,j)= C_{ref}$, and averaging~(\ref{eq:IwithAlpha}) to get
\begin{equation}\label{eq:getAlpha}
	\alpha(i,j) = \frac{\bar{I}(i,j)-I_0(i,j)}{C_{ref}}
\end{equation}
where the overbar denotes the average over $N$ images (in practice, we used $N=100$). 

\begin{figure}[t!]
	\centerline{
		\includegraphics{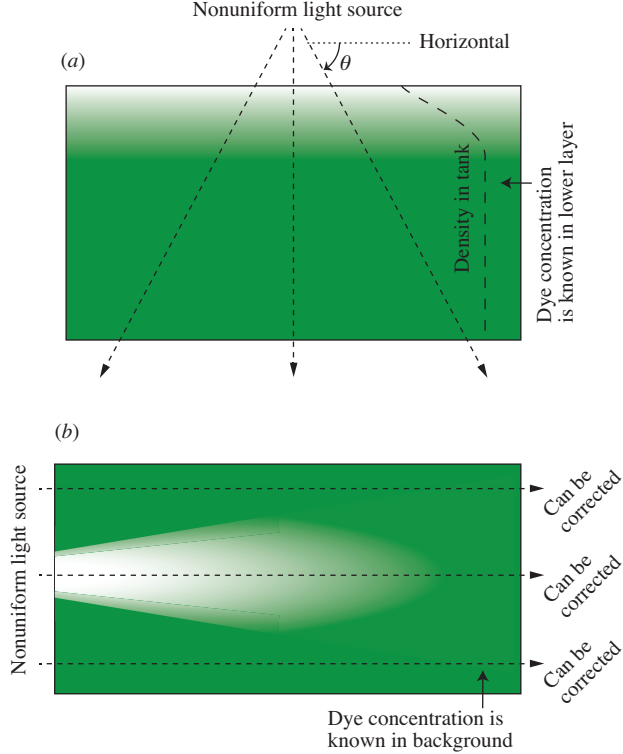}}
	\caption{($a$) Sketch of the power correction approach used in our experiment: we ensure that the bottom layer is at a known concentration, thereby enabling us to correct the complete field of view. ($b$) A possible extension to problems involving jet flows (as sketched previously in figure~\ref{fig:GuillardSketch}); here the ambient fluid is dyed, whereas the jet is clear (or at a different known concentration). Low maximum concentrations must be used to minimize attenuation (the light path could also be shortened by making it vertical in this figure, so as to further minimize attenuation).}
	\label{fig:rays}
\end{figure}

As explained in the introduction, in order to correct for power fluctuations, we need to ensure that each ray traverses a region with known dye concentration. In our problem, the deeper portions of the tank remain at constant concentration throughout the experiments. For this reason, we choose to set up the experiments such that the maximum dye concentration is reached in the bottom layer. In practice, we achieve this by dyeing the saltwater reservoir, and leaving the freshwater reservoir clear. Therefore, in the bottom layer, the concentration is known, that is $C(i,j) = C_{bot}$, as sketched in figure~\ref{fig:rays}($a$) (for completeness, a sketch of the corresponding implementation for the case of a jet shown earlier in figure~\ref{fig:GuillardSketch} is sketched in figure~\ref{fig:rays}$b$). Therefore, for each image $n$, we can compute, along each ray (characterized by azimuthal angle $\theta$):
\begin{equation}\label{eq:a}
	a_n(\theta)  = \frac{I_n(i,j)-I_0(i,j)}{\alpha(i,j) \; C_{bot}}, \quad \quad (i,j)\text{\;in\;bottom\;layer.}
\end{equation}
Once $a_n(\theta)$ has been computed, we can therefore estimate $C_n(i,j)$ as
\begin{equation}
	C_n(i,j)  = \frac{I_n(i,j)-I_0(i,j)}{\alpha(i,j) \; a_n(\theta)}.
\end{equation}

\subsection{Correcting the effect of PIV particles on measured fluorescence, while minimizing streaks\label{sec:scatter}}
In the absence of stratification, one might sometimes assume that particles are uniformly distributed, and simply perform the calibration (\ref{eq:getAlpha}) in the presence of particles. However, in stratified flows, particles will migrate toward their neutral buoyancy depth, making this technique not applicable.
One solution, mentioned in the introduction, involves deriving calibration curves as a function of particle and dye concentrations \cite{Law_Wang_ETFS_2000}. However, in order to use this technique one needs to estimate the particle concentration, which varies across the field.
We suggest a simpler alternative, which is well-suited to the stratified flows considered here. The key assumption behind the technique is that both the PIV particles, as well as the dye, behave approximately as Lagrangian tracers (that is, any diffusion taking place between the calibration and the experiment is negligible). Our procedure is constructed as follows.
\begin{figure*}[t!] 
	\centerline{
		\includegraphics{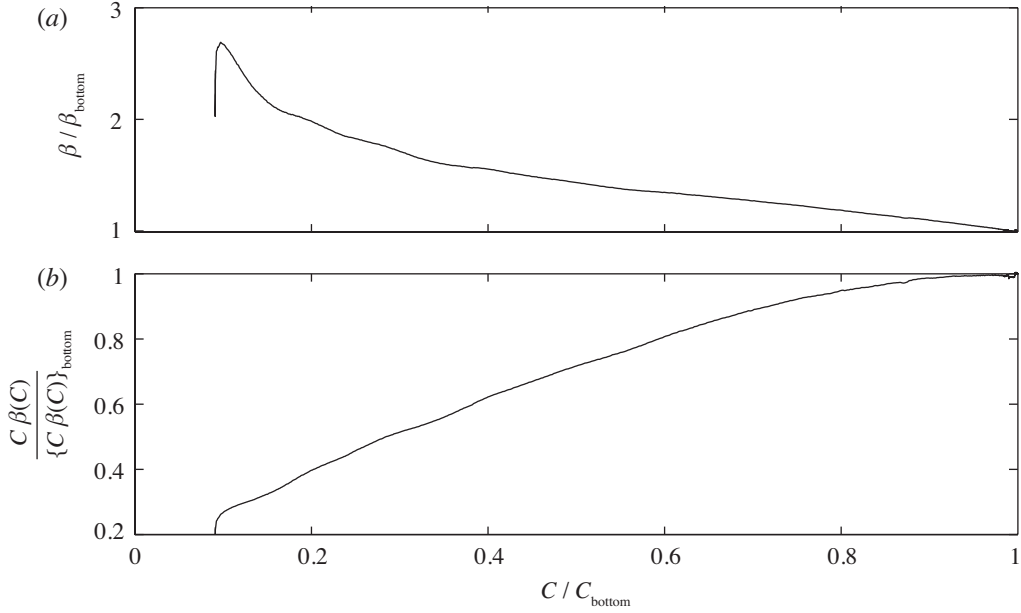}}
	\caption{($a$) An example showing $\beta$, the illumination enhancement due to a local excess of PIV particles, as a function of dye concentration $C$, illustrating the fact that $\beta$ can be well-approximated by a function of $C$ in these problems. ($b$) Relationship between $C\; \beta(C)$ and $C$, showing that the mapping is invertible, but becomes singular as one approaches $C = C_{bot}$ (since $d[C\;\beta(C)]/dC \rightarrow 0$ here). Note that the flat slope at large $C$ is due to the fact that $C$ is uniform in the lower layer, in our experiment.}
	\label{fig:beta}
\end{figure*}

\begin{figure*}
	\centering
	\includegraphics[width=0.8\linewidth]{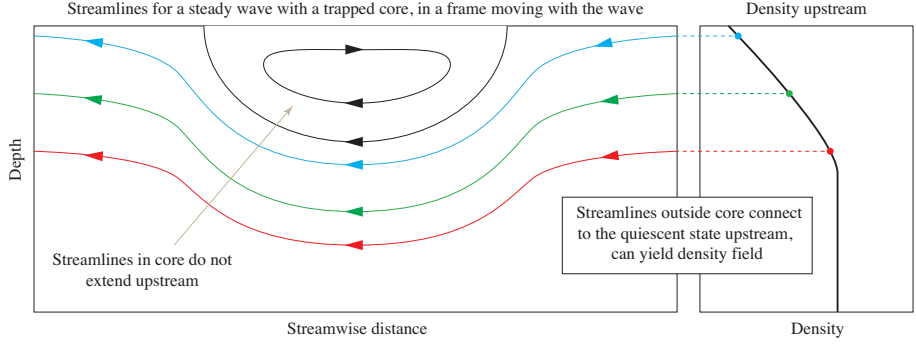}
	\caption{Schematic illustration for the procedure used to continue the upstream density profile (which is known) into the wave region, for steady waves. Since for steady flow isopycnals coincide with streamlines, the streamfunction calculated from PIV data is used for this purpose. This procedure cannot be applied to obtain the density in the core interior, where streamlines are disconnected from the upstream region.}
	\label{fig:densityFromStreamlines}
\end{figure*}

Firstly, we calculate $\alpha(i,j)$, using a solution of uniform concentration, without any PIV particles, using (\ref{eq:getAlpha}). We assume that the relationship between measured fluorescence and dye concentration can be written as
\begin{equation}\label{eq:IwithBeta}
	I_n(i,j) = \alpha(i,j) \, a_n(\theta) \, \beta_n(i,j) \, C_n(i,j) + I_0(i,j),
\end{equation}
where $\beta_n$ depends on the local particle concentration. We now describe a procedure to calculate $\beta_n$. Once the tank has been filled, before running an experiment, we measure the density profile $\rho_{init}(z)$ (where $z$ is the vertical direction) using a conductivity-temperature probe. Since, by construction, $\rho$ and $C$ are linearly related, we have $C_{init}(z)$ from
\begin{equation} 
C_{init}(z) = \frac{\rho_{init}(z) - \rho_{bot}}{\rho_{top}-\rho_{bot}}(C_{top}-C_{bot}) + C_{bot}.
\end{equation}

In this initial state, we may reasonably assume that the PIV particle concentration is only a function of $z$ (since particle migration is due to buoyancy, and occurs vertically), such that $\beta_{init}(i,j) = \beta_{init}(z)$. Furthermore, since in the stratified layer $C_{init}$ changes monotonically with $z$, we may invert the $C_{init}(z)$ relation and write $z = z_{init}(C)$, which in turn enables us to write, in this region:
\begin{equation} \label{eq:betaDef}
	\beta(i,j) = \beta_{init}{(z)} = \beta_{init}{(C)},
\end{equation}
such that, at this initial time, $\beta$ can be calculated as (after averaging equation~\ref{eq:IwithBeta} over $N$ images, with the fluid at rest)
\begin{equation} \label{eq:betaInit}
	\beta_{init}(C_{init}) = \frac{\bar{I}(i,j)-I_0(i,j)}{\alpha(i,j) \; C_{init}(z)}.
\end{equation}
Note that this requires that $C>0$ everywhere in the field of view, which is easily achieved by appropriately dyeing the reservoirs (in our experiments, $C$ was small but finite at the surface).

At this stage we make use of the key assumption introduced earlier in this section, namely that PIV particles and dye will both behave as Lagrangian tracers. Since $\beta$ is affected only by the local particle concentration, this implies that (\ref{eq:betaDef}) will remain applicable throughout the evolution of the flow, such that $\beta{(i,j)} = \beta_{init}(C_{init}) \equiv \beta(C)$ {\it for all times}.

At each instant in time, we now calculate the power correction using a slightly modified version of (\ref{eq:a}), which now includes $\beta$
\begin{equation}
	a_n(\theta)  = \frac{ I_n(i,j)-I_0(i,j) }{\alpha(i,j) \; \beta(C_{bot}) \; C_{bot}},\quad (i,j)\text{\;in\;bottom\;layer.}
\end{equation}
We can use this to calculate the product of $C(i,j)$ and $\beta( C(i,j) )$:
\begin{equation} \label{eq:betaC}
	\beta( C_n(i,j) ) \; C_n(i,j)  = \frac{ I_n(i,j)-I_0(i,j) }{ \alpha(i,j) \;a_n(\theta) }, 
\end{equation}
where all quantities on the right-hand-side are known. 

A somewhat technical (but practically important) requirement to extract $C(i,j)$ is that the mapping relating $\beta(C) \, C$ and $C$ must be invertible (that is, we need $d[\beta(C)\,C]/dC \neq 0$). Note that $\beta$ corresponds, essentially, to an amplification factor associated with the presence of particles; if these are neutrally buoyant at a level inside the stratified layer, $\beta$ will have a maximum inside the layer. Therefore, to ensure invertibility, $C$ must change sufficiently rapidly across the layer.
Since $\beta(C)$ is known from (\ref{eq:betaInit}), we store the relation $\beta(C) \,C$ versus $C$ in a lookup table, and we readily extract $C$ from the values of $\beta(C) C$ found using (\ref{eq:betaC}). Since, in order to construct this mapping, the dye and particles must follow Lagrangian trajectories, we propose referring to this technique as a ``Lagrangian correction''. Finally, the density field $\rho_{strat}$  in the stratified layer is found from $C$ as
\begin{equation} \label{eq:C2rho}
	\rho_{strat}(C) = \frac{C - C_{bot}}{C_{top}-C_{bot}}(\rho_{top}-\rho_{bot}) + \rho_{bot}.
\end{equation}

An example from our data, showing $\beta$, as well as $\beta(C)\, C $ versus $C$, is shown in figure~\ref{fig:beta}. Note that, at the edge between the stratified region and the deep, homogeneous layer (corresponding to the maximum $C$), $\beta(C)\,C$ changes weakly with $C$, meaning that this relationship cannot be inverted reliably. This implies that our Lagrangian correction is not applicable in this transitional layer. To address this issue, we exploit the fact that the background density profile is known, and that the flow will revert to the background (initial) state sufficiently far ahead of the wave. Furthermore, at the edge of the stratification, and further below (away from the bottom boundary layer) the flow is essentially two-dimensional (regardless of any instabilities in the stratified layer), and streamlines can be computed reliably from PIV in our experiments (as assessed by comparing streamlines obtained using different paths of integration).

\begin{figure*}[t!]
	\centerline{
		\includegraphics{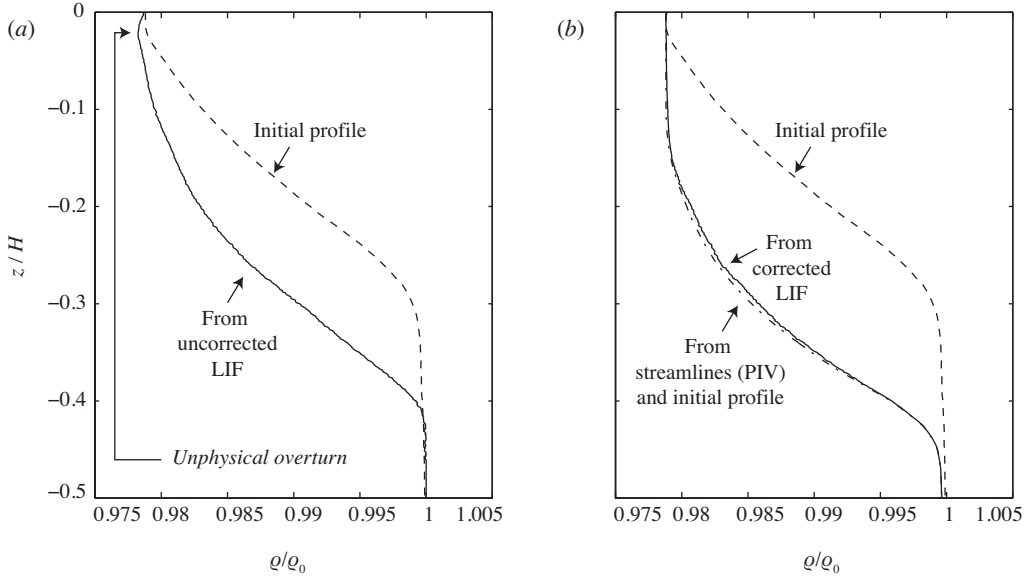}}
	\caption{Example showing results before and after our correction has been applied. ($a$): the dashed line shows the initial (background) density profile, measured using the conductivity-temperature probe. The continuous line corresponds to the density profile at the wave trough, calculated from the LIF data, without applying a correction for the presence of PIV particles. An unphysical overturn is present near the surface. ($b$) illustrates the results after applying our correction (continuous line). Since this wave is stable and does not have a trapped core, in this example we can use streamlines (from PIV) together with the initial profile to separately calculate the density at the wave trough (dot-dashed line), thereby validating our Lagrangian correction.}
	\label{fig:densProfiles}
\end{figure*}

\begin{figure*}
	\centering
	\includegraphics[width=0.8\linewidth]{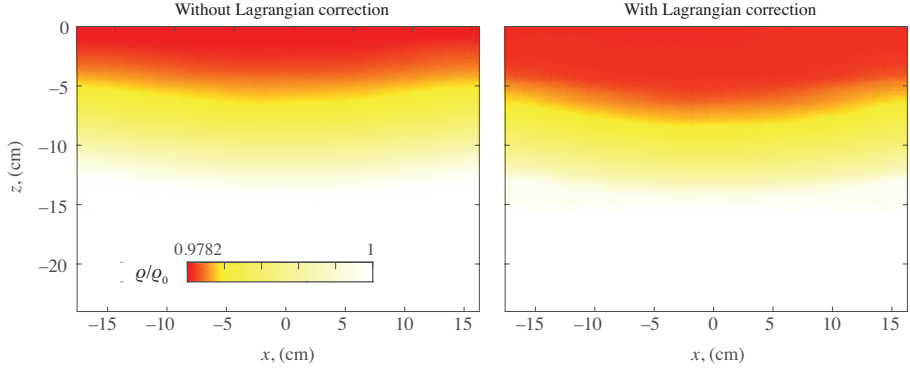}
	\caption{Density fields calculated without (left) and with (right) application of the Lagrangian correction. This field corresponds to the trough measurements shown earlier in figure~\ref{fig:densProfiles}. Each field is an average of three images.}
	\label{fig:densityFieldsBeforeAndAfterBetaCorr}
\end{figure*}

If the flow in this bottom region is near-steady, then streamlines (in the frame of reference moving with the wave) will coincide with contours of constant density (note that, if this were not the case, there would be a velocity component normal to the density contours, which would result in a time-varying shape, in the frame of reference translating with the wave). Therefore the density field in the deep region, starting below the stratified layer can be found as
\begin{equation} 
	\rho_{deep}(x,z) = \rho_{deep}(\psi(x,z)) = \rho_{init}(\psi_{init}(z)),
\end{equation}
where $\psi$ is the streamfunction in the frame translating with a wave with phase velocity $c$, such that it is related to the fixed-frame streamfunction by $\psi = \psi_{fixed}-c z$, and $\psi_{init}(z) = -c z$. In practice, we use $\psi_{init}$ and $\rho_{init}$ to construct a table that maps $\psi$ to $\rho$, then for each value of $\psi(x,z)$ in the flow field we use this table to interpolate the corresponding $\rho(x,z)$. A sketch of this procedure is shown in figure~\ref{fig:densityFromStreamlines}.

Finally, we combine $\rho_{deep}$ and $\rho_{strat}$ using a weight-function $W$, such that
\begin{eqnarray} 
	\rho(x,z) &=& [1-W(\rho_{strat}(x,z)) ] \; \rho_{strat}(x,z) \nonumber \\ 
	&+& W(\rho_{strat}(x,z))  \; \rho_{deep}(x,z),
\end{eqnarray}
where we set
\begin{equation} 
	W(\rho_{strat}) = \frac{1}{2} \left\{1+\text{erf}\left(\frac{\rho_{strat}-\rho_{switch}}{\rho_{scale}} \right) \right\},
\end{equation}
where we choose $\rho_{scale} = 0.1 (\rho_{bot}-\rho_{top})$ and $\rho_{switch} = 0.993\, \rho_{bot}$. 
An additional benefit of inferring density from PIV data, in this deep region, is that any streaks in the LIF data (which would originate with instabilities in the stratified layer) do not affect the deep layer density field. Therefore this approach also helps to mitigate some index-of-refraction effects.
\begin{figure*} 
	\centerline{
		\includegraphics[width = 0.75\textwidth]{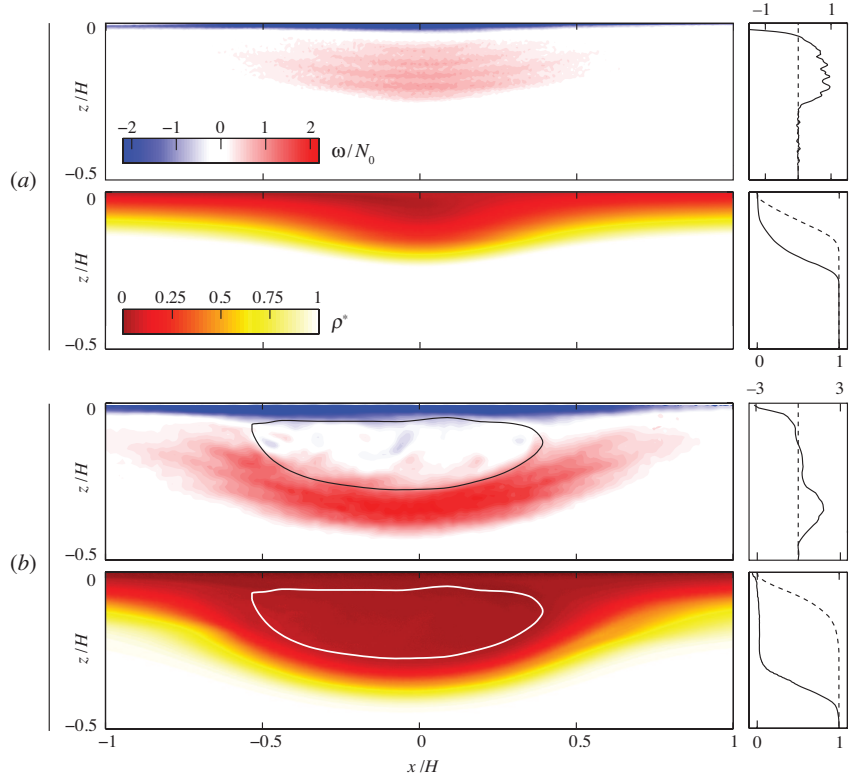}}
	\caption{Vorticity (upper panel of pair) and density (lower panel of pair) fields
		for waves without and with trapped cores (from \cite{Luzzatto-Fegiz_Helfrich_JFM_2014}). Core regions are drawn in ($b$). Right-hand panels show vertical slices at the wave trough; background conditions are marked by dashed lines. Note that these waves propagate from left to right.}
	\label{fig:waves}
\end{figure*}

An example illustrating the various steps described above is shown in figure~\ref{fig:densProfiles}. Here, for the sake of testing the procedure, we choose a small-amplitude wave, whose velocity field is essentially two-dimensional everywhere.
Panel ($a$) shows the initial density profile (dashed line), as well as the estimated density profile at the wave trough, calculated using a LIF procedure that neglects the effect of PIV particles (which corresponds to setting $\beta = 1$ everywhere). Note that, near the surface, this would erroneously indicate an overturn, with a density value below the lowest density introduced in the system! In panel ($b$), the density profile estimated by our technique, featuring the Lagrangian correction, is shown by the black continuous line. Note that no overturns are present. Figure~\ref{fig:densityFieldsBeforeAndAfterBetaCorr} shows the corresponding density fields, calculated without (left) and with (right) the Lagrangian correction. The two results are clearly markedly different.

Since this wave is essentially two-dimensional, in this specific example $\rho$ can be found reliably from the PIV at all depths. The corresponding density profile is shown by the dot-dashed line in figure~\ref{fig:densProfiles}($b$). This compares very favorably with the profile found using our Lagrangian correction on the LIF data, thereby providing a check on the accuracy of our procedure. However, our corrected LIF technique is also applicable to flows where the density may not be inferred exclusively from the PIV data, as shown in the next section.

\section{Examples involving regular and trapped-core waves\label{sec:example}}
As a further illustration of our technique, here we briefly show two examples, involving both regular and trapped-core waves (the flow physics associated with these results are discussed in detail in \cite{Luzzatto-Fegiz_Helfrich_JFM_2014}). Figure~\ref{fig:waves}($a$) shows a smaller-amplitude solitary wave. The top-left panel corresponds to the vorticity $\omega = \partial u/\partial z - \partial w/ \partial x$, whereas the panel below shows the density field (where the rescaled density $\rho^*$ ranges between zero and one). The right-hand panels show profiles of vorticity and density at the wave trough; dashed lines indicate the initial profiles.

Figure~\ref{fig:waves}($b$) displays similar data to figure~\ref{fig:waves}($a$), but for a trapped-core wave. The closed contours, in the left-hand panel, show the trapped core region (this is identified as a ``Lagrangian coherent structure'', by examining the rate at which neighboring particles separate \cite{Luzzatto-Fegiz_Helfrich_JFM_2014}). As highlighted by the right-hand panels, we discover that the trapped-core region has an approximately uniform density (and vorticity), thereby providing essential information for the development of oceanographic models of these waves.
In further work, we have used the data obtained through our combined PIV/LIF technique to also evaluate other wave properties, including kinetic and potential energy, as well as to test stability theories for stratified flows \cite{Luzzatto-Fegiz_Helfrich_JFM_2014}.

\section{Conclusions \label{sec:conclusions}}
In this paper, we address three issues affecting simultaneous PIV and LIF in stratified flows. These issues comprise (1) time- and space-dependent variations in the laser sheet; (2) preferential clustering of PIV particles near their neutral buoyancy depth, leading to locally higher dye illumination and fluorescence; and (3) index-of-refraction variations, affecting both illumination and data acquisition. Our experiments consist of large-amplitude solitary internal waves, propagating in a stratification consisting of a thin approximately linearly stratified region overlaying a deep uniform-density layer.

Issue (1) is addressed by exploiting the fact that, in our experiments, the dye concentration does not change in the deeper layer, thereby allowing us to calculate an instantaneous power correction, for each ray, using a reference fluorescence (based on a long-term average) from the lower layer. In order to deal with issue (2), we devise an additional correction technique, based on the assumption that, in the stratified layer, both PIV particles and fluorescent dye behave in a purely Lagrangian manner. In the deeper layer, we use streamlines from the PIV (together with the initial density profile) to calculate the density field. This last step, together with a setup that minimizes the optical path in the tank, also helps to contain the effects of index-of-refraction variations.

We test the overall technique by considering a small-amplitude, nonbreaking wave example, where the density field can also be obtained solely by using streamlines from PIV, thereby providing a direct comparison for the density extracted from the LIF. The two data sets are in good agreement. Finally, we show vorticity and density fields for larger amplitude waves, including a wave with a trapped core, where density could not be obtained exclusively using streamlines from PIV. Our technique has provided, to the best of our knowledge, the first experimental measurements of density fields for long internal waves with trapped cores. A full discussion of these results is presented in a separate paper \cite{Luzzatto-Fegiz_Helfrich_JFM_2014}.

We should clarify that some of the correction steps presented here are not always applicable; for example, if diffusion of dye is significant during the experiment (as may be the case, for example, in a flow dominated by turbulence), the Lagrangian assumption will not hold, and the correction for the presence of PIV particles will not be feasible. In those cases, one may pursue an alternative approach, such as simultaneous PIV and synthetic schlieren (see \cite{Dalziel_etal_MST_2007}; note that the schlieren measurements implicitly report a density average across the lateral extent of the tank, rather than being coplanar with the PIV).

However, it is worth noting that the corrections introduced here require no additional investment or complications in the experimental setup. In fact, our approach eliminates the need for a power meter (which would, in any case, not enable an accurate power correction across the laser sheet), and the additional effort associated with our corrections is applied almost exclusively at the postprocessing stage. We expect that the approach described here may prove valuable in other problems involving stratified flows.\\[6pt]

We gratefully acknowledge support by the Postdoctoral Scholar Program at the Woods Hole Oceanographic Institution, the Devonshire Foundation, and Churchill College, Cambridge. We are grateful to Anders Jensen for his work on the construction of the wave tank, and for his assistance with running the experiments.


\end{document}